\documentclass[prl,nobibnotes,floatfix,superscriptaddress,twocolumn,showpacs]{revtex4-1}
\usepackage{amsfonts,amsmath,graphicx,epsfig,latexsym}
\usepackage{subfigure}
\usepackage{graphicx}
\usepackage{color}
\usepackage{natbib}
\usepackage{multirow}
\usepackage{latexsym,amssymb}
\usepackage{amsmath}
\usepackage{soul}

\begin{document}

\title {How correlated is the FeSe/SrTiO$_3$ system? }
\author{Subhasish Mandal}
\affiliation
{Department of Applied Physics, Yale University, New Haven, Connecticut 06511, USA}
\author{Peng Zhang}
\affiliation  
{Department of Physics, Xi'An Jiaotong University, Xi'an, Shaanxi 710049, P.R. China}
\author{Sohrab Ismail-Beigi}
\affiliation
{Department of Applied Physics, Yale University, New Haven, Connecticut 06511, USA}
\author{K. Haule}
\affiliation  
{Department of Physics, Rutgers University, Piscataway, New Jersey 08854, USA}

\begin{abstract}
{\footnotesize  
Recent observation of $\sim$ 10 times higher critical temperature in FeSe monolayer compared with its bulk phase has drawn 
 a great deal of attention because the electronic structure in the monolayer phase appears to be different than bulk FeSe. Using a combination of density functional theory and dynamical mean field theory,
we find electronic correlations have important effects on the predicted atomic-scale geometry and the electronic structure of the monolayer FeSe on SrTiO$_3$.  
The electronic correlations are dominantly controlled by the Se-Fe-Se angle either in the bulk phase or the monolayer phase.
But the angle sensitivity increases and the orbital differentiation decreases in the monolayer phase compared to the bulk phase. 
The correlations are more dependent on Hund's J than Hubbard U. The observed orbital selective incoherence to coherence crossover with temperature confirms 
the Hund's metallic nature of the monolayer FeSe. We also find electron doping by oxygen vacancies in SrTiO$_3$ increases the correlation strength,
especially in the $d_{xy}$ orbital by reducing the Se-Fe-Se angle.}
\end{abstract}


\maketitle
\newpage

{\it Introduction:} In addition to the cuprates, the discovery of superconductivity in Fe-based compounds with superconducting critical temperatures (T$_c$) 
ranging from 26 K to 56 K has created a new class of unconventional superconductors \cite{doi:10.1021/ja800073m,Mazin:2010he,iron1}. Recent observations of T$_c$ 
reaching as high as 100K in FeSe monolayer grown on SrTiO${_3}$ (STO) have further boosted interest to search for high T$_c$ superconductors in this 
family \cite{STO1,STO2,STO3,STO4,PhysRevLett.115.017002,EP4,ncomms1,R_Peng}.  
 Photoelectron spectroscopy measurements show that, unlike other Fe-based superconductors, the Fermi surface of single-layer (one unit-cell)  
 FeSe on STO consists only of electron pockets at the zone corners (X-point), without the hole pockets around the zone center ($\Gamma$-point) \cite{ARPES1,ARPES2,ARPES3,R_Peng}. 
 This can lead to a different mechanism of gap opening other than the sign changing s-wave pairing state from spin fluctuation found in Fe-based superconductors in 
 its bulk phase \cite{PhysRevLett.101.057003}. 
 Apart from the Fermi surface, there are many contrasting signatures observed in the monolayer phase of FeSe when compared to the bulk pnictides. For example, the FeSe/STO 
 system was suggested to be in close proximity to a Mott-insulating phase where an insulator-superconductor crossover was found. It was then concluded that similar to the 
 cuprates, the correlations strength was found to be controlled by the Hubbard-U interaction \cite{ARPES3,ncomms1}. 
A recent study based on the density functional theory (DFT)  shows antibonding hybridization between Fe-{\it d} and Se-{\it p} increases in the monolayer and thus can 
lead to decreased electron correlation through increasing bandwidth \cite{PhysRevLett.115.017002}. 
 Various experiments also suggest electron doping makes the FeSe/STO system more correlated \cite{e-dope1,ncomms1,ARPES2}, opposite to what usually happens in the bulk 
 iron pnictides \cite{Ba122}. More contrasting behavior is noticed where the enhancement in T$_{c}$ in FeSe/STO was suggested to arise from strong interfacial
 effect \cite{ARPES2,ARPES1,R_Peng}.

\begin{figure*}
\includegraphics[width=500pt, angle=0]{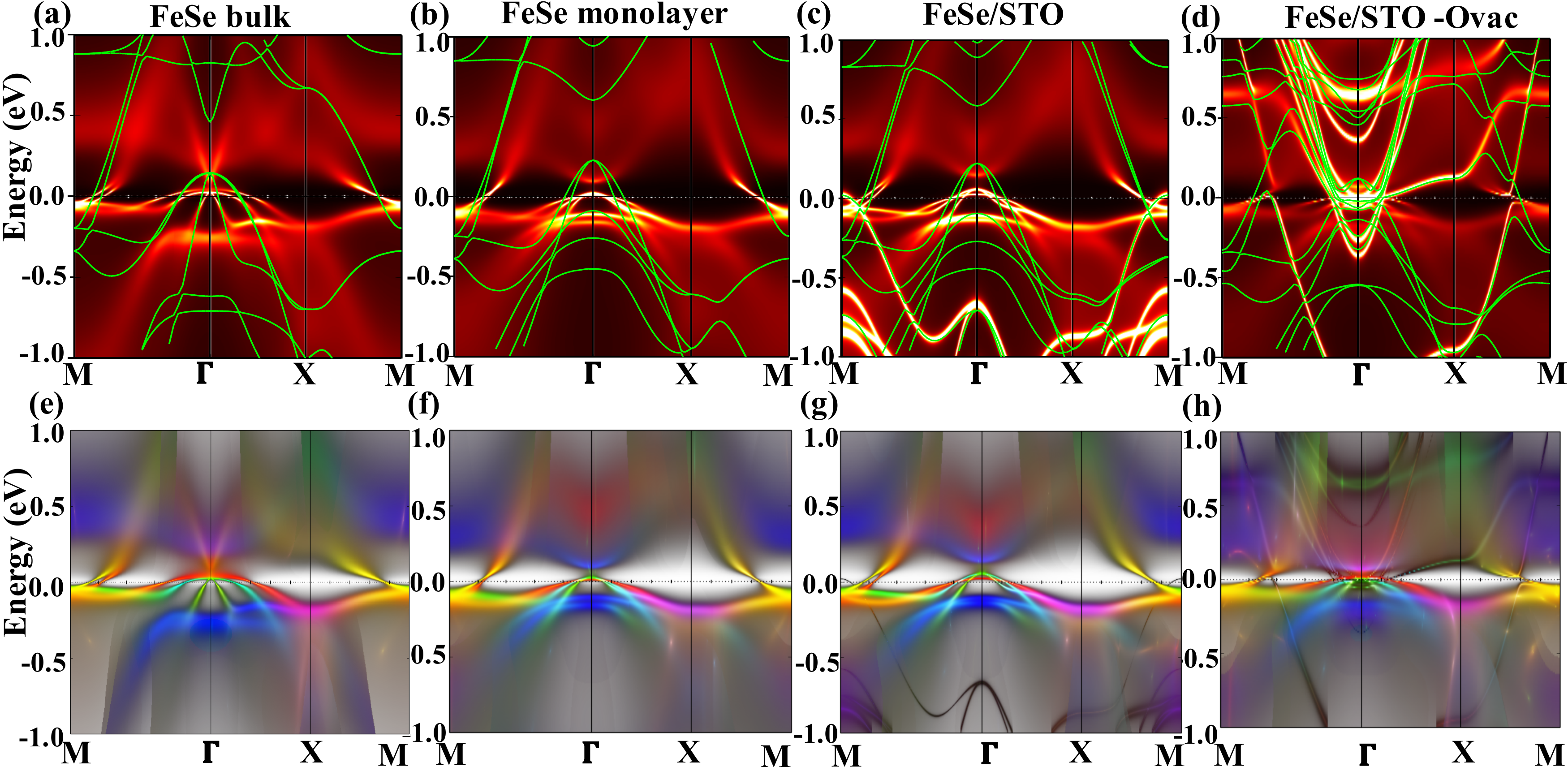}
\caption{(Color online).
Top (a-d):Computed DFT+DMFT spectral function together with band structures computed within DFT (green lines) 
for FeSe bulk (a), FeSe monolayer(b), FeSe/STO (c) and FeSTO-Ovac (d). 
Down(e-h): Corresponding orbital resolved DFT+DMFT spectral functions; 
d$_{z^2}$ and d$_{x^2-y^2}$ are in blue, d$_{xz}$ and d$_{yz}$ are in green, d$_{xy}$ is red. 
}
\end{figure*}

While there are many density functional based studies describing the role of electron-phonon \cite{EP1, EP4,PhysRevB.93.134513}, 
band structure \cite{PhysRevB.87.220503,PhysRevB.92.035144,PhysRevB.93.180506,PhysRevLett.115.017002}, 
and the epitaxial growth of FeSe/STO \cite{PhysRevB.87.220503,PhysRevB.93.180506}, so far there is no direct explanation of why FeSe/STO behaves 
so dramatically different than its bulk phase \cite{doi:10.1063/1.4965892} using a method
that truly captures fluctuating local moments. Using DFT in combination with dynamical mean field theory (DFT+DMFT) method \cite{RevModPhys.78.865,PhysRevB.81.195107} 
we study four phases of FeSe and STO:
(I) FeSe bulk, 
(II) a freestanding FeSe monolayer, 
(III) FeSe on a SrTiO$_3$ substrate without oxygen vacancies (FeSe/STO) and 
(IV) FeSe on SrTiO$_3$ substrate with 50\% oxygen vacancies (FeSe/STO-Ovac) since oxygen vacancies can be a potential source of electron doping as observed in recent experiments 
and theory \cite{O_vac,ARPES2,PhysRevB.87.220503,PhysRevB.92.035144,PhysRevB.93.180506}. In addition we study a freestanding FeSe monolayer, chopped from FeSe/STO-Ovac structure 
maintaining the same Se-Fe-Se angle of FeSe/STO-Ovac to investigate the effect of substrate. 
Here we attempt to address the following questions: i) How does the electron correlations change from the bulk phase to the monolayer phase? 
Typically going from bulk to two dimensions, electron correlations increase due to the reduction of electron's hopping in one direction. Is it true for Fe-pnictide? 
ii) What is the origin of electron correlations? Hubbard-U or Hund's coupling-J or both? 
iii) Can electron correlations change the topology of the Fermi surface compared to that obtained in conventional DFT? 
iv) Why electron doping through O vacancies in STO increases the correlation in the FeSe monolayer?

{\it Method and structural details:} The pcnitide height and/or bond angle between X-Fe-X (X=pnictide) plays an important role in determining the strength of 
the correlations \cite{PhysRevLett.112.217202,PhysRevB.86.060412} and T$_c$ \cite{PhysRevB.81.205119} across various compounds in the bulk phase. 
However its experimental determination for the monolayer phase is still on going. 
An accurate determination of the structural parameter is essential due to its extreme sensitivity in controlling the strength of correlation and T$_{c}$. 
In this letter, relaxed structures are obtained using the self-consistent DFT with embedded DMFT method that incorporates the effect of the electron's entropy 
while computing forces on atoms. 
The implementation of the force optimization within DFT+DMFT and the resulting accuracy in obtaining the pnictogen height in bulk FeSe is described in Ref\cite{PhysRevB.94.195146}. 
To compare the effect of spin-fluctuation and electron entropy in the structural optimization, we also obtain 
atom positions from both nonmagnetic (NM) and spin-polarized (SP) flavors of DFT.
More details of our methods and structural information are described in the SI \cite{SI}.  In Table I in the SI \cite{SI}, we describe the key structural parameters. 
The DFT+DMFT-computed bond angle of Se-Fe-Se in FeSe/STO and the pncitide height (h$_{Se}$) are closer to the spin-polarized DFT than the nonmagnetic DFT \cite{Linscheid} 
--indicating the local spin-fluctuation and the long range order have a similar effect in determining h$_{Se}$ and the angle \cite{PhysRevB.94.195146,EP5}. 
However DFT+DMFT predicts structures with smaller Se-Fe-Se angle than SP-DFT. Going from bulk FeSe to monolayer, the Se-Fe-Se angle increases from 103$^\circ$ to 109$^\circ$ and
h$_{Se}$ is decreased by 7.0\% (See SI \cite{SI}). With introduction of oxygen vacancies in STO, the angle reduces to $\sim$ 107$^\circ$.
Interestingly the angle in the monolayer phase is close to the "magic" angle where the T$_c$ is found to be the highest in the bulk phase of Fe-pnictides \cite{PhysRevB.81.205119}. 

\begin{figure}
\includegraphics[width=220pt, angle=0]{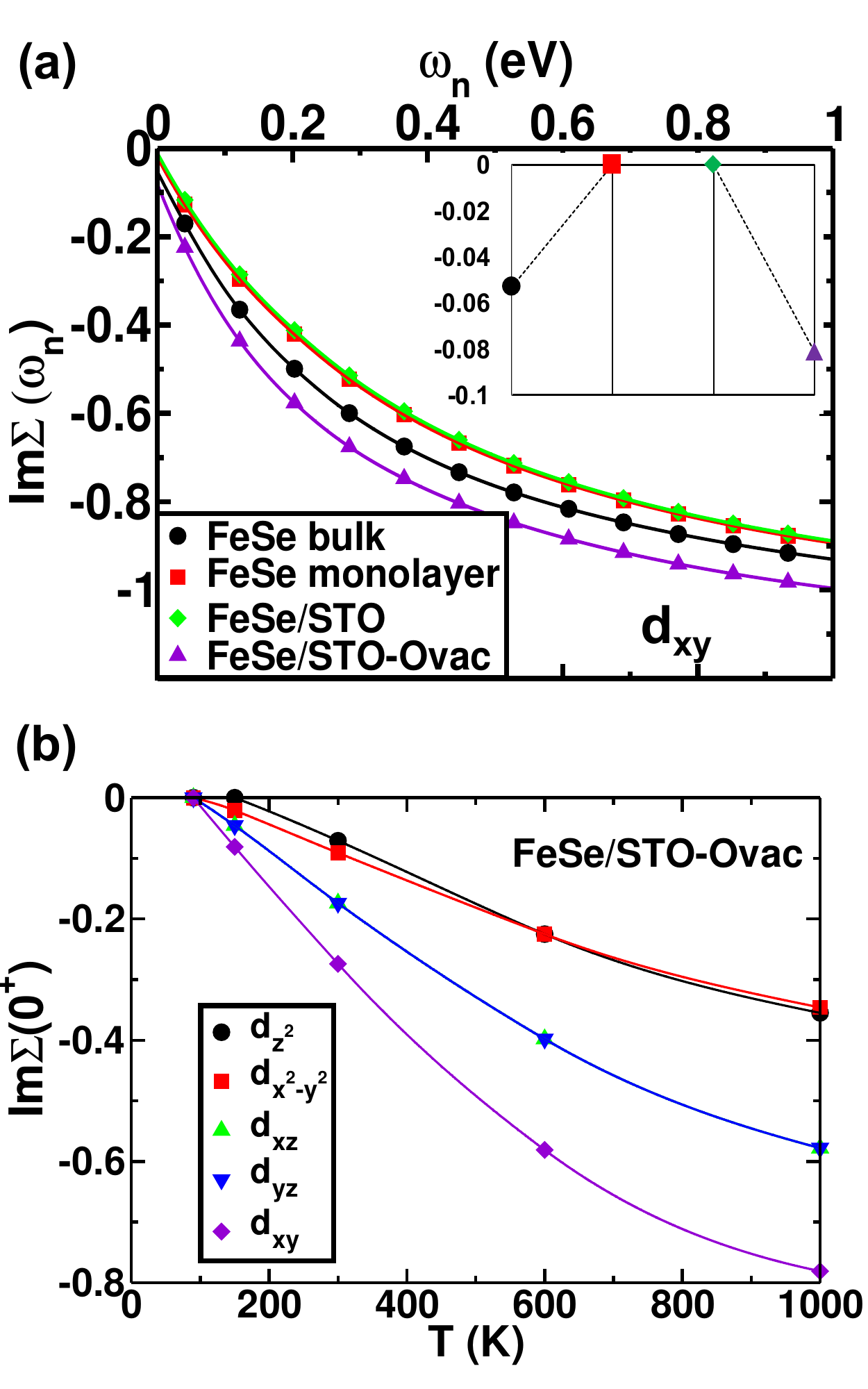}
\caption{(Color online).
 (a) Imaginary part of self-energy $Im\Sigma(i\omega_n)$ of the Fe-d$_{xy}$ orbital for four different systems;
inset shows the extrapolated value Im $\Sigma(i\omega_n \rightarrow i0^+)$ reflecting structure dependent 
coherence to incoherence crossover at 150K.  Subfigure (b) shows temperature dependence of Im$\Sigma(i\omega_n \rightarrow i0^+)$ in FeSe/STO-Ovac 
for the Fe-3d orbitals to show temperature driven coherence-incoherence crossover with orbital dependent crossover temperatures. 
}
\end{figure}

{\it  Spectral function:} Once we optimize the atom positions using DFT+DMFT, we compute spectral functions shown in Figs. 1(a) -1(h). 
Understanding the strength of electron correlations and the nature of the Fermi surfaces are fundamental tasks to understand unconventional superconductivity \cite{RevModPhys.75.473,PhysRevLett.79.3506,PhysRevLett.87.177007}. 
On the top panels of Figs. 1(a)-1(d), we show DFT+DMFT computed  spectral functions on the same color scale for four different systems. 
The green lines indicate the DFT band structures computed for the SP DFT-optimized structures in the monolayer phase. 
Brighter color in the spectral function reflects quenched correlation. 
From bulk to monolayer phases, we notice a significant shrinking in the size of the hole pockets around $\Gamma$. 
This is not prominent in the DFT bands, which reflects the effect of correlation to be important. 
The increase in the sharpness of the DFT+DMFT spectral function is noticed  while going from the bulk phase to the FeSe-monolayer and 
hints that they exhibit a different degree of electron correlation.  
From bulk to monolayer phases, the DMFT spectral function becomes more coherent -- indicating the suppression of correlations in the FeSe monolayer and FeSe/STO. 
With O vacancies,  the spectral function around $\Gamma$ again becomes dim, which reflects the increase of correlation.  
We find significant changes in the topology of the Fermi surface in the monolayer phase, especially around the $\Gamma$-point. 
The spectral function in freestanding FeSe monolayer and FeSe/STO are very similar. But introducing O vacancies in STO makes the Fermi surface significantly different.

To identify the orbital dependent nature of electron correlation effect, we compute DFT+DMFT orbital-dependent spectral functions for the four systems and plot them together on 
the bottom panel in Fig 1(e-h). The electron  and hole pockets are mainly made of  d$_{xy}$ (red) and d$_{xz+yz}$ orbitals (green). 
Going from bulk to the monolayer phases, the xy-pocket around $\Gamma$ shrinks the most and also changes the sharpness and becomes less correlated in FeSe monolayer 
and FeSe/STO and again  becomes correlated in FeSe/STO-Ovac. 
With the introduction of O vacancies in STO, the DMFT spectral function becomes more correlated for  all t$_{2g}$ orbitals. 
Computed DFT+DMFT spectra show that the hole pocket at $\Gamma$ point shrinks significantly and almost vanishes when O vacancies are introduced while the electron pockets 
are found to get bigger.  This leads to electron doping similar to Refs\cite{PhysRevB.87.220503,PhysRevB.92.035144,PhysRevB.93.180506}.  

\begin{figure}
\includegraphics[width=220pt, angle=0]{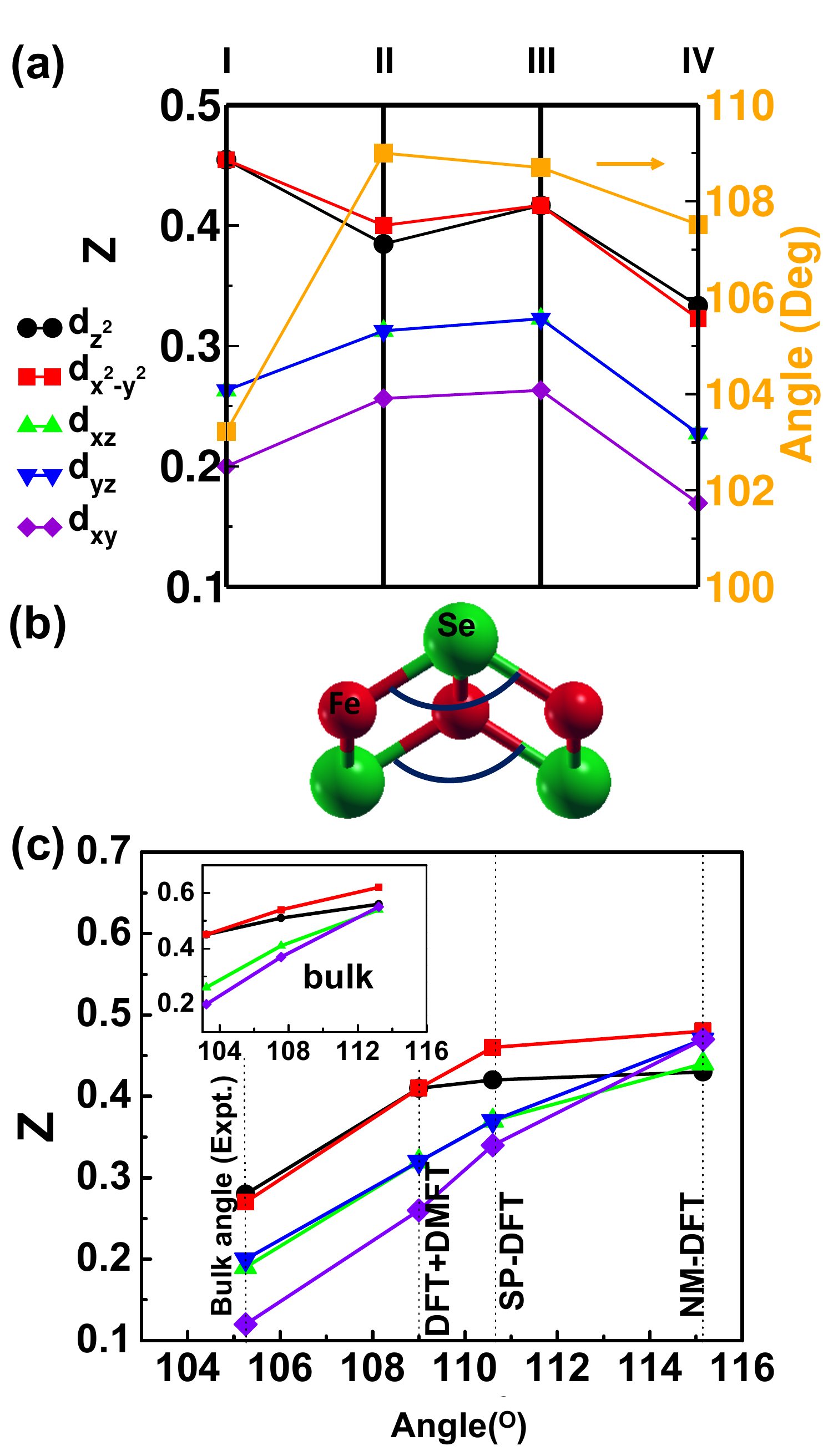}
\caption{(Color online).
(a) The spectral weight (Z) on different Fe-3d orbitals (left) and Se-Fe-Se angle (right) show a similar trend across four structures: 
I) FeSe bulk, II) free-standing FeSe monolayer, III) FeSe/STO, and IV)FeSe/STO-Ovac.  
Subfigure (b) is a schematic representation of Se-Fe-Se angle in FeSe. The angle dependent spectral weight on different Fe-3d orbitals are presented in 
(c) for free-sanding FeSe monolayer. The dashed lines represent the Se-Fe-Se angle obtained from experiment and optimized with NM-DFT, SP-DFT, and DFT+DMFT methods. 
Inset represents the angle dependence of Z for bulk FeSe.   
}
\end{figure}

{\it Coherent scale:} 
In Fermi liquid theory, the inverse quasiparticle lifetime equals to the scattering rate $\Gamma$=$-ZIm\Sigma(i0^+)$, 
where $Z$=$(1-\partial Re\Sigma(\omega)/\partial \omega)|_{\omega\rightarrow0}$ is the spectral weight
and $Im\Sigma(i0^+)$ is the imaginary part of self-energy at zero frequency. At low temperatures when
$Im\Sigma(i0^+) \rightarrow 0$, the system is in the coherent phase with infinite quasiparticle lifetime. 
When the temperature is above the coherent energy scale, $Im\Sigma(i0^+)$ and, consequently, the quasiparticle lifetimes 
are both finite. Our DFT+DMFT calculations show that the coherent scales of Fe-3d electrons are strongly orbital dependent 
(see SI \cite{SI} for all five {\it d} orbitals)  and also tuned by the structure of the FeSe/STO system. 
In Fig. 2a, the imaginary part of quasiparticle self-energy Im$\Sigma(i\omega_n)$ of the Fe-3d$_{xy}$  orbital at 150 K 
are extrapolated to $i0^+$.  The extrapolated values are shown in the inset. This directly shows the coherent scales for the four different systems. 
  Im$\Sigma(i0^+)$ of FeSe-bulk and FeSe/STO-Ovac are sizable (Fig. 1a and SI-Fig. S1) at 150 K,
which indicates that the d$_{xy}$ orbital is incoherent and the coherent scales of FeSe/STO-Ovac  and bulk FeSe are similar.
In contrast, although Im$\Sigma(i0^+)$ of the FeSe monolayer and FeSe/STO are still finite, 
their absolute values are much smaller indicating their coherence at 150 K.  
The behavior of Im$\Sigma(i0^+)$ for these four systems at a fixed temperature indicates that the monolayer structure of FeSe and its growth on STO 
substrate greatly enhance their corresponding coherent 
energy scales in t$_{2g}$ orbitals. Dramatically, oxygen vacancies in STO again change the coherent scale in all Fe-{\it d} orbitals.  

Since the Fermi surface of FeSe/STO-Ovac is the closest to the ARPES measurements, we examine the oxygen vacant FeSe/STO system in more detail. 
First, we investigate the effect of U and J on $Z$ in FeSe/STO-Ovac.
 When J is changed from 0.8 to 0.5 eV with a fixed U, the change in Z is found to be stronger than when U is changed from 5.0 eV to 2.0 eV with a fixed J
 (Table II in SI \cite{SI}). This shows the system to be more sensitive to change in J than U and also confirms the Hund's metallic nature found in bulk pnictides.
We also explore the temperature driven coherence-incoherence crossover in FeSe/STO-Ovac. 
Although it is well known that bulk pnictides are incoherent bad metals \cite{hauleC,Yin:2011ca,PhysRevB.84.224520,Nekrasov2013}, 
there is no study found in the literature for the monolayer phase.  
A recent study based on the slave-Boson approach showed that the $d_{xy}$-orbital of FeSe/STO behaves like a Mott insulator \cite{ncomms1}. 
We show the temperature dependence of the extrapolated values Im$\Sigma(i0^+)$ of FeSe/STO-Ovac in Fig.2b. 
At 90 K, all Im$\Sigma(i0^+)$ are very small, proving FeSe/STO-Ovac is in the coherent state. As temperatures increase from 90 K to 1000 K, 
Im$\Sigma(i0^+)$ move away from zero to finite values, which signatures a temperature driven coherence-incoherence crossover in FeSe/STO-Ovac. 
The d$_{xy}$ orbital shows the most temperature dependence.

{\it Quasiparticle weight:} The structural-tuned coherent scales are directly related to the electron correlations.
To examine degree of electron correlations in more detail, we compute the orbital dependent spectral weights $Z$
(inverse of mass-enhancement $m^*/m_{band}$) after analytic continuation of the self-energy by the maximum entropy method \cite{Jarrell1996}.
 $Z$ is unity in a noncorrelated system, and goes to zero in the strongly correlated limit. 
We compute $Z$ for all the Fe-3d orbitals of the four different structures (Fig. 3a).  
First,  we notice that the correlation in bulk FeSe has much more orbital differentiation than in the monolayer phase. 
We find that the d$_{xy}$-orbital is the most kinetically frustrated for both bulk \cite{haule3} and monolayer phases. 
Going from bulk to monolayer, all t$_{2g}$ orbitals become less correlated and all e$_g$ orbitals become more correlated.  
Introducing oxygen vacancies in STO increases correlation in all orbitals. This effect is extremely sensitive in the  d$_{xy}$-orbital, which 
is the most correlated orbital; Z becomes even smaller than in the bulk phase. In FeSe/STO-Ovac, Z in d$_{xy}$ orbital almost halves indicating 
the very strong correlations as observed in the experiment \cite{O_vac,ncomms1}.
Here an obvious question arises: why O vacancies make FeSe/STO so strongly correlated? 
To answer that, we show the  Z as a function of the Se-Fe-Se angle for the four systems (Fig. 3a).  
Z for all t$_{2g}$  orbitals follow a similar pattern as the Se-Fe-Se angle increases: 
they first increase from bulk to monolayer and then decrease with the introduction of O vacancies.  

In the monolayer phase, the decrease in Z with O vacancies directly relates to the decrease in angle.  
To prove this, we compute Z as a function of Se-Fe-Se angle in FeSe-monolayers. Fig. 3c shows a monotonic behavior as we increase the angle.  

Three angles are obtained from NM-DFT, SP-DFT, and DFT+DMFT optimization of FeSe monolayers. The fourth angle is  for a monolayer structure with Se-Fe-Se angle of bulk.  
This directly shows that a monolayer structure with the angle of bulk FeSe is extremely correlated and the correlation is controlled by one single structural parameter, 
which is the Se-Fe-Se angle.  We also plot the similar angle dependence of Z in bulk FeSe (Fig. 3c inset). 
In bulk FeSe, the sensitivity of angle in Z is much less than that in monolayer, especially for the $d_{x^2-y^2}$ and the $d_{z^2}$ orbitals. 

{\it Role of oxygen vacancies:} O vacancies serve as a potential source for doping electrons to FeSe as seen in our DFT+DMFT spectral function 
as well as in the experiments \cite{O_vac,ARPES2,PhysRevB.87.220503,PhysRevB.92.035144,PhysRevB.93.180506,Dagdeviren2016}. To explore the effect of doping on Z, 
we construct a monolayer structure with the angle of FeSe/STO-Ovac after chopping out the substrate (FeSe-chopped).  
It is interesting to note that Z in FeSe-chopped is slightly lower compared to that of FeSe/STO-Ovac.  
For FeSe-chopped the Z is 0.31, 0.31, 0.22, 0.22, 0.17, while for FeSe/STO-Ovac they are 0.33 0.32  0.23  0.23  0.17 respectively for $d_{x^2-y^2}$, $d_{z^2}$, $d_{xz}$, $d_{yz}$, 
and $d_{xy}$ orbitals. However the change in Z is not large since a very small amount of charge transfer happens to the Fe-3d orbitals (only 0.02 electron charge-transfer is found).
It is well known that doping reduces correlation in the bulk pnictides due to weakening of Hund's rule coupling \cite{Ba122}.  
Our results support this and directly show that it is the reduced angle with O vacancies that increases the correlation.  
This explains the experimental observation and puzzle of increasing correlation with electron doping in FeSe/STO system via O vacancies. 
In summary, we show two fold effects of O vacancies: they dope electrons to the FeSe and they increase the correlation because they decrease the Se-Fe-Se angle.

{\it DFT+DMFT hybridization:} The change in the Se-Fe-Se angle can affect the hybridization (or bandwidth). The DFT+DMFT hybridization is shown in S.I. Fig. 2 \cite{SI}. 
We notice that the hybridization around the Fermi energy in all $t_{2g}$ orbitals follow a pattern: starting from bulk, it increases for the freestanding FeSe monolayer 
and FeSe/STO and then again decreases for the FeSe/STO-Ovac. This shows the effect of correlations on the monolayer is directly related to the hybridization 
modulated by the Se-Fe-Se angle.

{\it Conclusions:} Our results show that the strength of correlations in FeSe/STO heterostructures is dominantly controlled by the Se-Fe-Se angle, 
which is sensitive to the oxygen vacancies in the STO. The quenched correlations in the free-standing FeSe monolayer and FeSe/STO are directly 
related to the increased hybridization due to the increase in the Se-Fe-Se angle. Introducing O vacancies in the STO reduces the angle and reduces the hybridization, 
as a result the system becomes more strongly correlated. Despite several reports claiming superconductivity in FeSe/STO 
to be mediated by electron-phonon coupling \cite{ARPES1,EP1,EP2,EP3}, 
 the strength of electron-phonon coupling in conventional DFT is found to be too low to explain the high T$_c$ \cite{EP2,PhysRevB.93.134513}. 
The structural-tuned increased electron correlations with oxygen vacancies in STO can enhance the electron-phonon coupling in FeSe/STO similarly to bulk FeSe \cite{EP5}. 
Our study favors unconventional superconductivity in FeSe/STO, likely with the orbital antiphase s$_{+-}$ pairing symmetry \cite{spin_haule}, 
where the two electron pockets have opposite sign of pairing, as the absence of hole pockets disfavors conventional s$_{+-}$ symmetry. \\

{\it Acknowledgements:}  We acknowledge R. E. Cohen and A. B. Georgescu for helpful discussions.
We acknowledge Mark Jarrell and Juana Moreno for their important help.
S. M. and P. Z. acknowledge the Carnegie Institution for Science.
S. M. and S. I. B acknowledge support from the NSF SI2-SSI program (Grant No. ACI-1339804).
S. I. B acknowledges partial support from NSF MRSEC DMR-1119826. 
P. Z. acknowledges the supports from NSFC-11604255.
K. H. acknowledges the supports from NSF DMR-1405303. 
Major computations are performed at the `Supermike' in Louisiana State University.  
Additional computations are performed at the NERSC supercomputing facility, Carnegie Institution for Science 
and Yale University Faculty of Arts and Sciences High Performance Computing Center.

\bibliography{super}

\pagebreak
\widetext
\begin{center}
\textbf{\large Supplemental Materials: How correlated is the FeSe/SrTiO$_3$ system ? }
\end{center}
{ \it Structural details and Methods} To investigate the effect of interface on the electronic structure of FeSe, we construct FeSe monolayer 
with and without SrTiO$_3$ (STO) substrate.  In our investigation we consider both undopped STO and STO with oxygen vacancies. 
We have optimized structures with  self-consistent DFT with Embedded DMFT \cite{PhysRevB.94.195146} 
for (I)FeSe bulk, (II) a freestanding FeSe monolayer, (III) FeSe on a SrTiO$_3$ substrate without oxygen vacancies (FeSe/STO) and (IV) FeSe on SrTiO$_3$ 
substrate with 50\% oxygen vacancies (FeSe/STO-Ovac). 
During the optimization of the slabs, the in-plane lattice parameter of the slab is fixed to the experimental lattice parameter of STO (3.905 $\AA$)  
since an epitaxial growth of FeSe was found in the experiment\cite{STO1,PhysRevB.93.180506}. 
Also a pristine FeSe monolayer chopped from the FeSe/STO-Ovac optimized structure is also constructed (without optimizing) 
to study the effect of O vacancies in STO while maintaining the same angle of FeSe/STO-Ovac. 
To compare the structures obtained from DFT+DMFT optimization,  both nonmagnetic and spin-polarized DFT calculations are also performed within the 
generalized gradient approximation of Perdew-Burke-Ernzerhof \cite{PhysRevLett.77.3865} for exchange and 
correlation using the plane wave pseudopotential approach as implemented in QUANTUM ESPRESSO with ultrasoft pseudopotentials\cite{PhysRevB.41.7892}. 
The plane wave kinetic energy cutoff is set to 30 Ry with a corresponding charge density cutoff of 300 Ry.  
The Brillouin zone is sampled by a uniform 6$\times$6$\times$1 mesh of k points per 1$\times$1 interfacial unit cell.  
The structural optimization is converged when all force components are less than 1.4 meV/atom. 
Similar structures were used in a separate study to investigate the role of double layer TiO$_2$ on STO with electron doping in FeSe \cite{PhysRevB.93.180506}. 
Two important structural parameters, pncitide height ($h_{Se}$) and Se-Fe-Se angle, of all four systems are described in Table I. 
In FeSe/STO-Ovac the DFT+DMFT computed angles are different in the two directions. We present both values in the table. 
In the main text the average value of the Se-Fe-Se angle for FeSe/STO-Ovac is used.

\begin{table*}
\centering
\begin{tabular}{ | c l c | c | c | c | c | c | c |  }
\hline
\multicolumn{2}{| c |}{System} & \multicolumn{2}{| c }{NM-DFT }  & \multicolumn{2}{| c }{SP-DFT} & \multicolumn{2}{| c |}{DFT+DMFT} \\ 
\hline
 &  &  h$_{Se}$ ($\AA$) & Angle ( $^\circ$)  &  h$_{Se}$  ($\AA$)  & Angle ( $^\circ$)  &   h$_{Se}$ ($\AA$)  & Angle ( $^\circ$)  \\
   \hline
   FeSe bulk  & &1.33$^\dagger$  &  109.78$^\dagger$  & 1.43$^\dagger$ & 105.41$^\dagger$ & 1.49&103.23\\
  \hline
   FeSe monolayer  & & 1.24  & 115.16 & 1.35 & 110.6 & 1.39 & 109.0  \\
   \hline
FeSe/STO  & & 1.24 &  115.38   & 1.35 & 110.7 & 1.40 & 108.7\\
  \hline
   FeSe/STO-Ovac  & & 1.23 & 115.07 & 1.35 & 110.6 & 1.42, 1.45 & 108.12, 106.92  \\
  \hline
   
\end{tabular}
\caption{\label{tab:5/tc}Two key structural parameters, pncitide height ($h_{Se}$) and Se-Fe-Se angle, for four systems computed by the nonmagnetic DFT (NM-DFT),
the spin-polarized DFT (SP-DFT) and the DFT+DFMT methods. For the monolayer phase of FeSe, the in-plane lattice is fixed at experimental value of STO (a=3.975 $\AA$) 
since an epitaxial growth was found in the experiment.   $^\dagger$ From Ref\cite{EP5}. }

\end{table*}


The DFT+DMFT \cite{RevModPhys.78.865,PhysRevB.81.195107} method can capture the local moments physics in paramagnetic FeSe and 
can successfully describe  fluctuating local moments and electron correlations \cite{haule_spin,Wang:2013gz}.   
In DFT+DMFT, the self-energy that samples all local skeleton Feynman diagrams is added to the DFT Kohn-Sham 
Hamiltonian \cite{RevModPhys.78.865,PhysRevB.81.195107}. This implementation is fully self-consistent \cite{PhysRevB.81.195107,haule3}. 
The iterations stop after full convergence of the charge density, the impurity level, the chemical 
potential, the self-energy, and the lattice and impurity Green's functions. The lattice is represented using the full potential linear 
augmented plane wave method, implemented in the Wien2k \cite{wien2k} 
(LDA-DFT). The continuous time quantum Monte Carlo method is used to solve the quantum impurity problem and to obtain the 
local self-energy due to the correlated Fe-3d orbitals.
The self-energy is analytically continued from the imaginary to real axis using an auxiliary Green's function.
The Coulomb interaction $U$ and Hund's coupling $J$  are fixed at 5.0 eV and 0.8 eV, respectively \cite{Kutepov:2010bu}. 
To study the effect of U and J, U=2 eV and J=0.5 eV are also used.  A fine k-point mesh of  at least 100 kpoints, 
and a total 100 million Monte Carlo steps for each iteration are used for the paramagnetic phase of the FeSe/STO at T=116K.
We also study FeSe/STO-Ovac at various temperatures to explore the coherence-incoherence crossover.  \\

{\it Results.} In S.I. Fig. 1 we show the structure dependent imaginary part of self-energy $Im\Sigma(i\omega_n)$ of Fe-3d electrons 
in (I) FeSe bulk, (II) free-standing FeSe monolayer, (III) FeSe/STO, and (IV) FeSe/STO-Ovac for (a) $d_{z^2}$, (b) $d_{x^2-y^2}$, (c) $d_{xz}$, (d) $d_{yz}$, (e) $d_{xy}$ orbitals. 
Subfigure (f) shows the Se-Fe-Se angle (right axis) controlled spectral weight (left axis) for four different systems.  In DMFT loop, the hybridization between the impurity and its mean-field bath is derived by 
$G_0^{-1}(i\omega)=i\omega+\mu-\Delta(i\omega)$, in which $\Delta(i\omega)$ is the hybridization function 
and $G_0^{-1}(i\omega)$ is the noninteracting Green's function. The hybridization functions of Fe-3d orbitals in (I) FeSe bulk, (II) free-standing FeSe monolayer, (III) FeSe/STO and (IV) FeSe/STO-Ovac in DMFT level 
are shown in S.I. Fig. 2. These functions show that the hybridization of Fe-3d electrons with other electrons in crystal generally follows 
the similar trends as the quasi-particle spectral weight (S.I. Fig. 1f) when we compare from bulk phase to monolayer phase.

 In Table II, we describe the orbital dependent spectral weight of FeSe/STO-Ovac system at various U and J at 116 K. This shows the system to be more sensitive  in J than U
and confirms the Hund's metallic nature.

\begin{center}
\begin{figure*}
\includegraphics[width=440pt, angle=0]{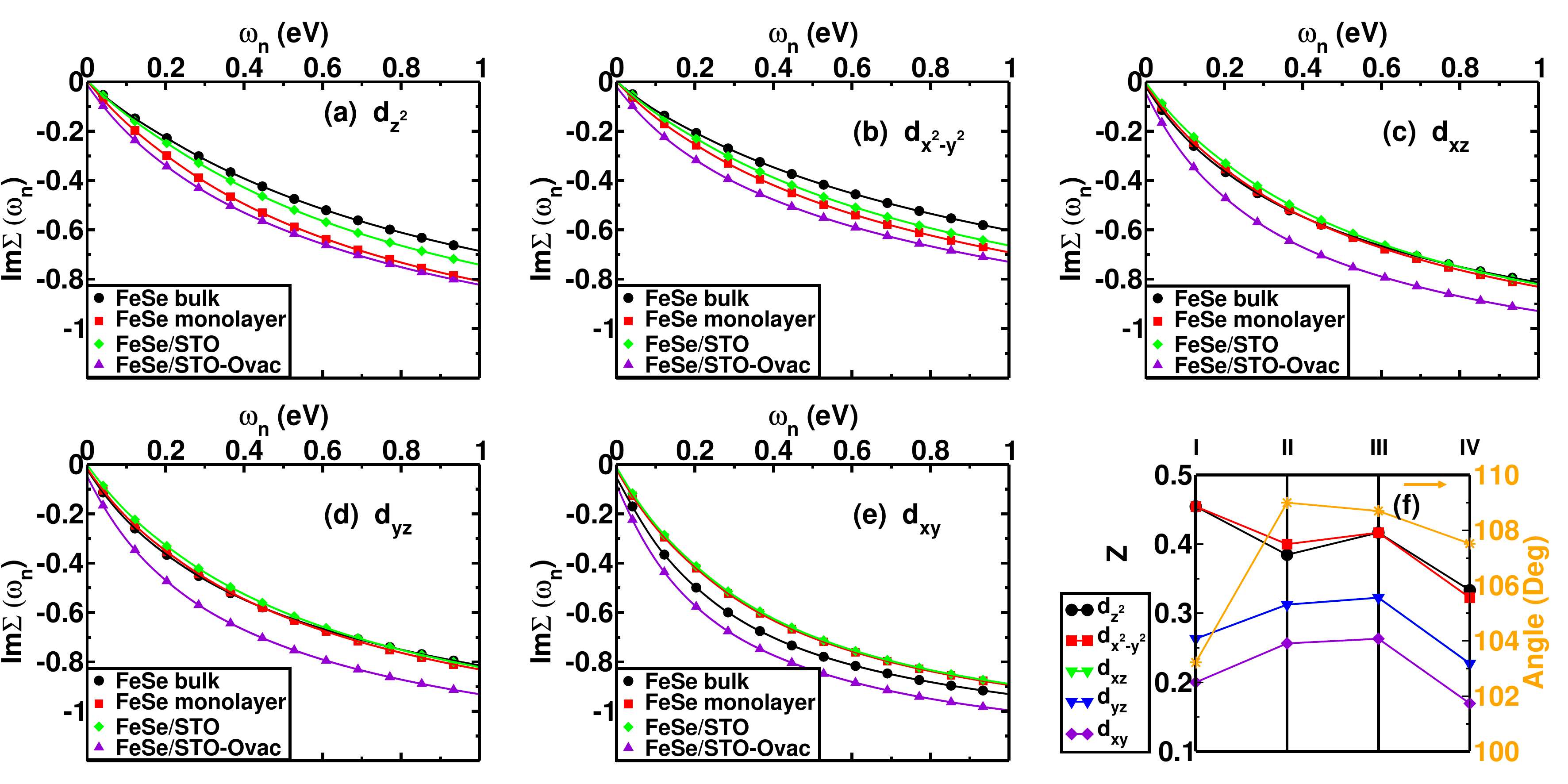}
\caption{ (Color online)
Structure dependent imaginary part of self-energy $Im\Sigma(i\omega_n)$ of Fe-3d electrons  for (a) $d_{z^2}$, 
 (b) $d_{x^2-y^2}$, (c) $d_{xz}$, (d) $d_{yz}$, (e) $d_{xy}$ orbitals. The solid symbols are 
 $Im\Sigma(i\omega_n)$ at Matsubara frequency and the solid lines are cubic spline of 
 $Im\Sigma(i\omega_n)$ to $i0^+$. Subfigure (f) shows Se-Fe-Se angle (right axis) controlled spectral weight (left axis)
for four different systems.  
 }
\end{figure*}
\end{center}

\newpage

\begin{center}
\begin{figure*}
\includegraphics[width=440pt, angle=0]{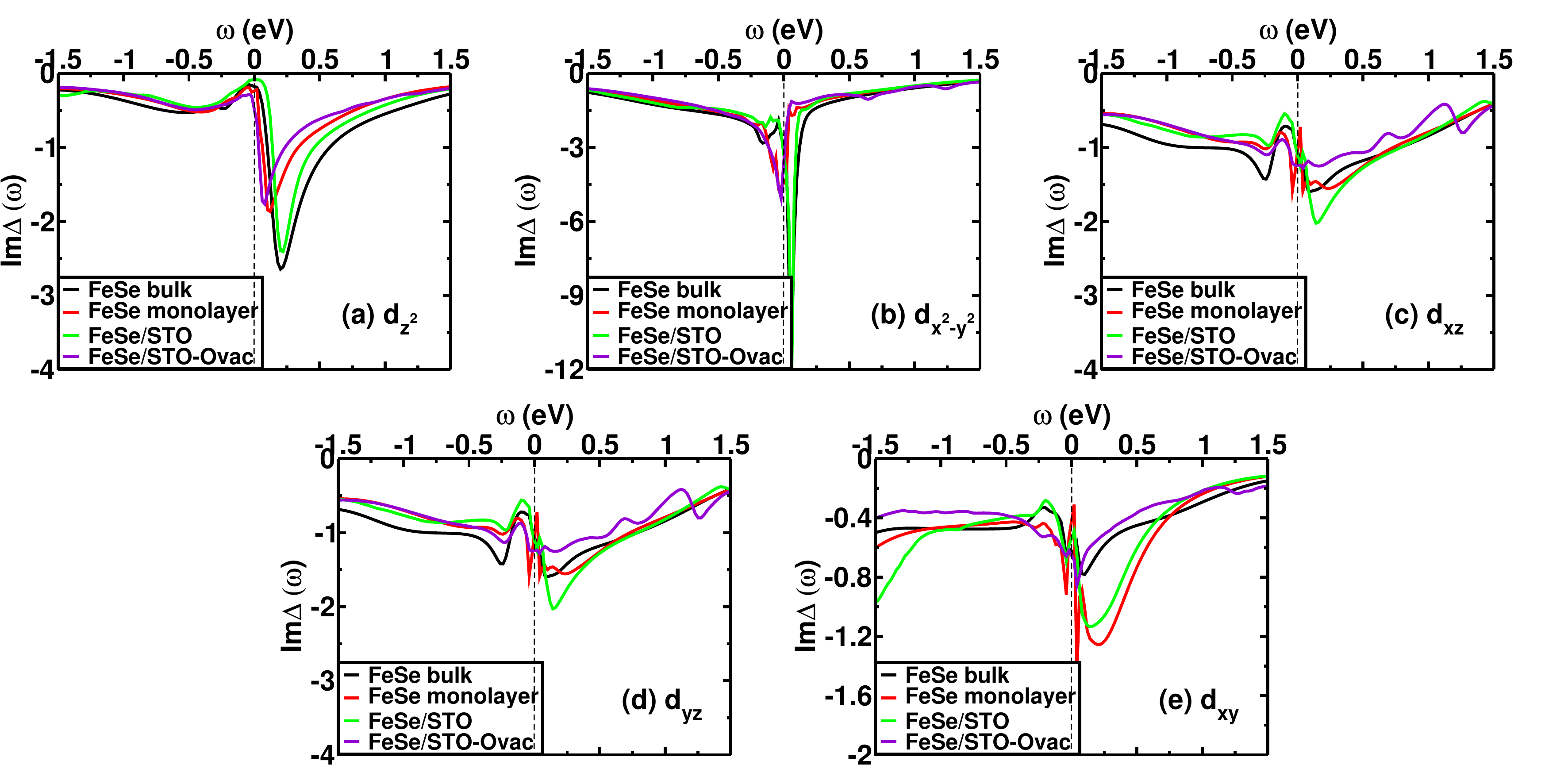}
\caption{ (Color online)
The hybridization functions of Fe-3d electrons in FeSe bulk, FeSe monolayer, FeSe/STO and FeSe/STO-Ovac using DFT+DMFT. 
 The hybridization functions are calculated using full self-energy correction within DMFT.
These functions show the hybridization follows a similar trend as the spectral weight in four systems for each orbital. %
  }
\end{figure*}
\end{center}



\begin{table}
\begin{tabular}{|c|c|c|c|c|c|c|c|} 
\hline
T(K)  & U (eV)   &J(eV)   &d$_{z^2}$  & d$_{x^2-y^2}$ & d$_{xz}$ & d$_{yz} $ &d$_{xy}$    \\
   \hline
   116 & 5.0 & 0.8 & 0.333 & 0.323 & 0.227 & 0.227 & 0.169 \\
   \hline
   116 & 5.0 & 0.5 & 0.572 & 0.539 & 0.499 & 0.489  & 0.472 \\
   \hline
   116 & 2.0 & 0.8 & 0.560 &  0.572 & 0.437 & 0.446  & 0.384 \\
   \hline
   
\end{tabular}
\caption{ Orbital dependent spectral weight of FeSe/STO-Ovac system at various U and J at 116 K.}

\end{table}


\end{document}